# 3D-PRINTING WATER-SOLUBLE CHANNELS FILLED WITH LIQUID METAL FOR RECYCLABLE AND CUTTABLE WIRELESS POWER SHEET

*Takashi Sato[1], Ryo Takahashi[2], Kento Yamagishi[2], Takao Someya[2], Michinao Hashimoto[3], Eiji Iwase[4], Yoshihiro Kawahara[2], Junya Kurumida[1], and Wataru Iwasaki[1]*

[1] National Institute of Advanced Industrial Science and Technology, JAPAN, [2]The University of Tokyo, JAPAN,
[3]Singapore University of Technology and Design, SINGAPORE, [4]Waseda University, JAPAN

## ABSTRACT

A recyclable and cuttable wireless power transfer (WPT) sheet is proposed, enabled by H-tree wiring and water-soluble channels filled with liquid metal (LM). Conventional 2D WPT systems lose their functionality when physically damaged or modified. The H-tree wiring pattern maintains the operation of the remaining coils even after the outer region of the sheet is cut away. The LM can be recovered by dissolving 3D-printed polyvinyl alcohol (PVA) channels in water. The sheet dimensions were experimentally optimized, and $Q$-factor over 55 was achieved at 6.78 MHz. The sheet maintained its bending stiffness (($2.54 \pm 0.10$) × $10^{-6}$ N·m$^2$) and electrical resistance ($7.6 \pm 0.3$ mΩ) during 100 bending cycles. After four dissolution–refabrication cycles, 98% of the LM was recovered with stable volume resistivity ($0.32 \pm 0.01$ mΩ·mm) and contact resistance ($1.7 \pm 1.3$ mΩ). The WPT sheet can be integrated into everyday objects and enables long-term, continuous operation of surrounding electronic devices, contributing to IoT society and ambient computing.

## KEYWORDS

Wireless power transfer, cuttable electronics, recyclable electronics, flexible electronics, liquid metal, galinstan, polyvinyl alcohol, H-tree wiring

## INTRODUCTION

In IoT society and ambient computing, numerous sensors and actuators are distributed throughout daily living environment and operated continuously [1]. Battery-powered devices require frequent charging and replacement [2], while wired-driven devices demand extensive cabling [3]. For these reasons, converting everyday objects, such as structures, furniture, and garments, into wireless power surfaces has attracted significant attention [4]. Flexible WPT sheets are particularly promising because they can be attached to a variety of existing objects [5]–[7].

Conventional 2D WPT systems require complex optimization of coil layout, mutual inductance, and matching conditions [5]-[7]. Designing WPT configuration for different shapes of items is impractical for non-experts. Ideally, users should be able to change the shape of WPT sheets and attach them to their target objects. However, conventional WPT sheets stop operating when physically damaged or modified. A cuttable WPT design that maintains transfer efficiency is needed.

WPT sheets also require recyclability because they may be disassembled and refabricated repeatedly for use on different objects. Commercial flexible printed circuit boards consist of copper–polyimide laminates, and their irreversible bonding makes layer separation difficult [8]. Silicone rubbers and conductive pastes are widely used in flexible electronics, but fabrication processes such as cross-linking and sintering are irreversible and hinder material recovery [9], [10]. Consequently, most WPT sheets with conventional flexible materials are discarded after use, and material recovery or reuse remains difficult for end users.

This paper proposes a recyclable and cuttable WPT sheet. The H-tree wiring pattern—previously demonstrated on copper–polyimide flexible substrates [11]—allows the remaining coils to operate even after the outer region is cut away. In this study, the pattern is integrated into water-soluble, LM–filled PVA channels to achieve both cuttability and recyclability. The LM can be recovered and reused by dissolving the PVA channels in water [12]–[15]. Based on the effect of channel thickness on both WPT performance and sheet flexibility, we experimentally determined the channel dimensions. Wireless powering of LEDs and repeated dissolution-refabrication tests confirmed stable operation and high recyclability of the sheet. The proposed sheet enables easy integration of WPT functions into various surfaces of everyday objects and supports continuous, long-term operation of electronic devices throughout the environment.

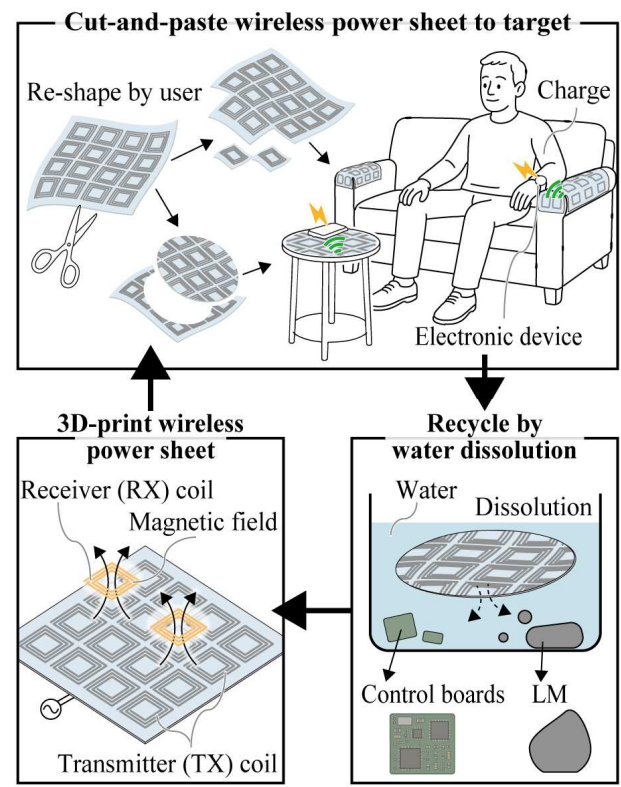

*Figure 1 Overview of recyclable and cuttable wireless power transfer (WPT) sheet by 3D-printed polyvinyl alcohol (PVA) channels filled with liquid metal (LM).*

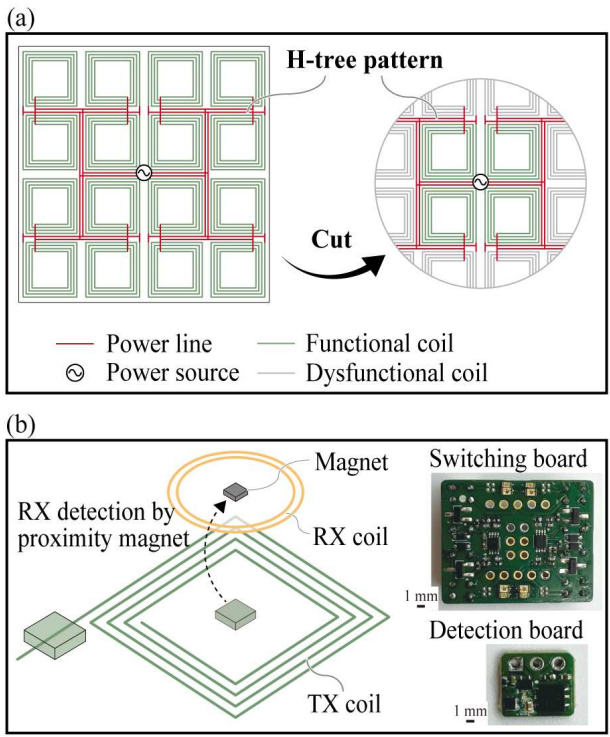

*Figure 2 Electrical design of the proposed sheet. (a) H-tree wiring that maintains power delivery after outer-edge cutting. (b) Selective power delivery enabled by RX-coil detection and switching circuits.*

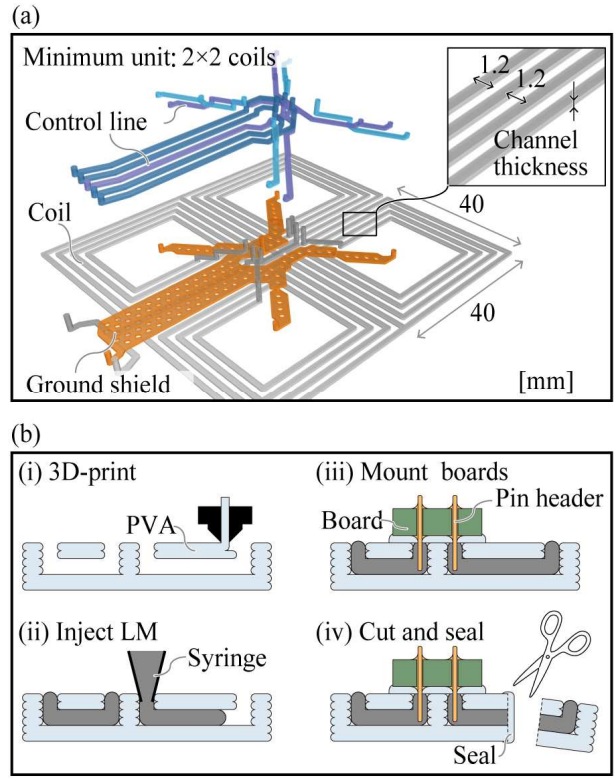

*Figure 3 Mechanical design of the proposed sheet. (a) Three-layer channel structure. (b) Fabrication process.*

## METHODS

The WPT sheets transfer power via magnetic coupling from a transmitter (TX) coil to a receiver (RX) coil. The H-tree wiring distributes power from the central module operating at 6.78 MHz to each coil with an equal path length (Figure 2a). The H-tree pattern can maintain the functionality of the remaining coils even after the outer region of the sheet is cut away. A Hall sensor-based detection circuit is placed at the center of each TX coil to sense the approach of an RX coil with a magnet (Figure 2b). The switching circuit prioritizes the power delivery to the TX coils near the RX coil.

The 3D-printed PVA (Bambu Lab) channel structure consists of three layers: coil, ground/shield, and control (Figure 3a). Each coil has four turns with a size of 40 × 40 mm. The channel width and spacing are 1.2 mm, and the channel thickness ranges from 0.24 to 4.8 mm. The inter-layer wall thickness is 0.48 mm. The overall sheet thickness is from 1.84 mm, with the mounted boards add 6.00 mm. The channels are fabricated using a fused deposition modeling (FDM) 3D printer (X1C, Bambu Lab) with a 0.4 mm nozzle, 0.08 mm layer height, and 100% infill. The pin headers of the detection and switching boards are inserted into the channel inlet/outlet to form electrical connection. The fabrication process is as follows: (i) 3D printing the channel structure (Figure 3b-i), (ii) injecting galinstan (Sichuan HPM) into the channels (Figure 3b-ii), (iii) inserting pin headers and sealing using a water-soluble adhesive (YAMATO) (Figure 3b-iii), and (iv) cutting the sheet using scissors and sealing the cut surface (Figure 3b-iv).

WPT efficiency, *i.e.*, Q-factor of a single coil, was measured using a vector network analyzer (VNA). Sheet cuttability was measured as the force required to cut through the channels with a blade. Flexibility was characterized by a three-point bending test with a curvature radius of 30 mm. The volume resistivity of the galinstan and the contact resistance between the galinstan and the pin headers were measured using a modified transfer length method.

## RESULTS AND DISCUSSION

Channel thickness was investigated as a key parameter determining both WPT efficiency and flexibility/cuttability of the sheet. Figure 4 shows the relationship between channel thickness and WPT performance. Galinstan could not be injected into channels thinner than 0.36 mm due to its high surface tension, and forced injection caused partial leakage. For thicknesses between 0.36 mm and 1.44 mm, the Q-factor increased monotonically and exceeded 55. Increasing the cross-sectional area of the coil suppressed the resistive losses. However, as the thickness increased beyond 1.44 mm, the Q-factor no longer improved and saturated at approximately 55–60. Increasing the channel thickness enlarged the conductor sidewall area and increased the stray capacitance between adjacent turns. As the stray capacitance increased, resistive loss no longer governed the $Q$-factor, and it saturated as a result. These results indicate that channels must be sufficiently thick to minimize resistive losses, but excessive thickness degrades the performance through stray capacitive coupling.

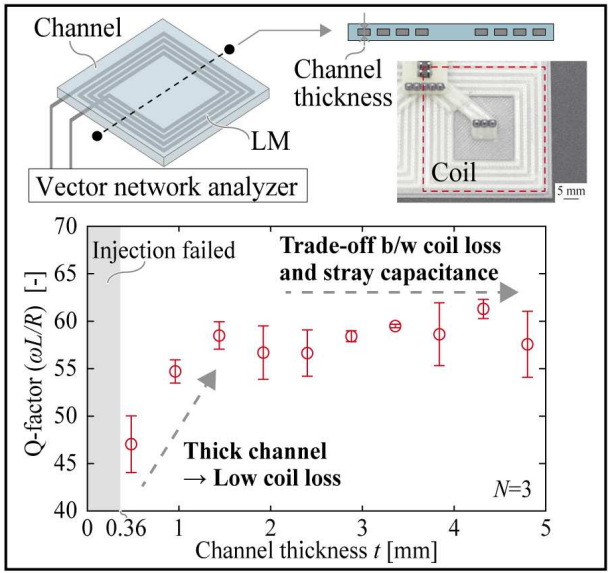

*Figure 4  Q-factor of a single coil for different channel thickness (N=3).*

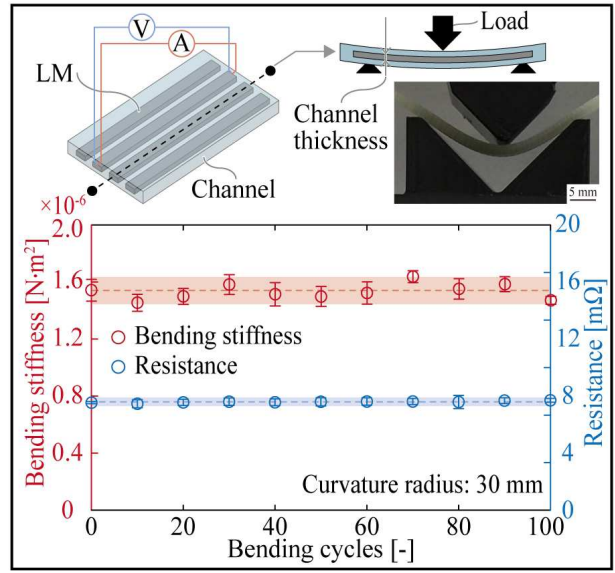

*Figure 6  Changes in bending stiffness and electrical resistance of the designed sheet during bending cycles (N=3).*

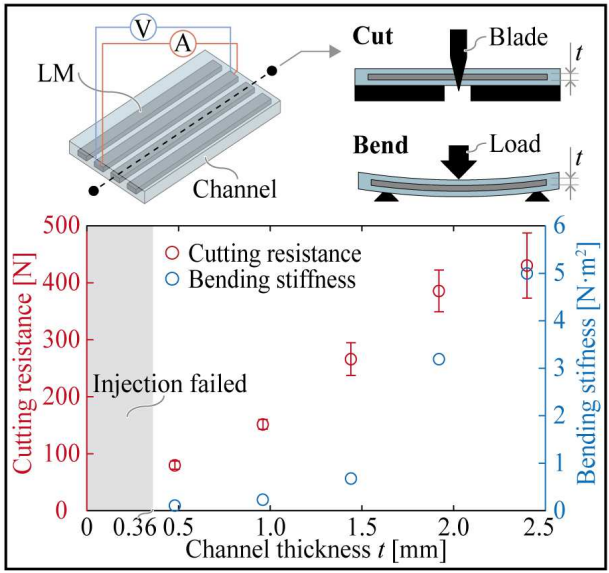

*Figure 5  Cutting resistance and bending stiffness of the sheet for different channel thickness (N=3).*

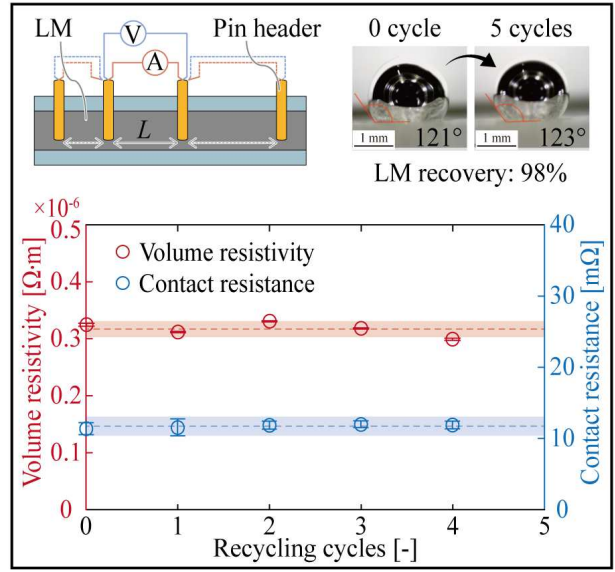

*Figure 7  Volume resistivity of recovered LM and contact resistance between LM and pin headers over four dissolution–refabrication cycles (N=1).*

Figure 5 shows the effect of the channel thickness on the cuttability and flexibility of the sheet. The thinner channels exhibited lower bending stiffness, enabling smooth bending of the sheet. The thinner channels also exhibited lower cutting force, making the sheet easier to reshape. However, for channels thicker than 1.92 mm, galinstan leaked from the cut surfaces during cutting. The surface tension was insufficient to hold the galinstan against its own weight, causing leakage. Therefore, thinner channels are required for both flexibility and cuttability. Based on this trade-off, the channel and sheet thicknesses were set to 1.44 mm and 2.40 mm, respectively. At these dimensions, the $Q$-factor had saturated while the sheet exhibited sufficient mechanical flexibility. Thus, the sheet with designed thicknesses enables the effective WPT without sacrificing flexibility or cuttability.

Next, the mechanical and electrical performance of the sheet was demonstrated. Figure 6 shows the cyclic bending durability of the sheet. During 100 bending cycles, the changes in bending stiffness and electrical resistance of the sheet were limited to $(2.54 \pm 0.10) \times 10^{-6}$ N·m$^2$ and $7.6 \pm 0.3$ mΩ, respectively. The sheet did not exhibit plastic deformation, crack formation, channel rupture, or inter-channel leakage. These results demonstrate that the sheet maintains stable electrical conductivity and mechanical flexibility under repeated bending. This durability is essential for integrating WPT sheets into flexible and deformable surfaces.

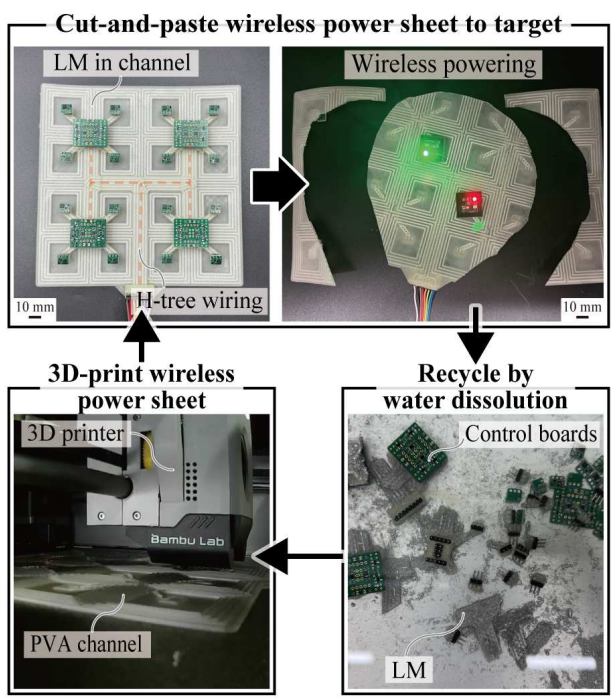

*Figure 8 Demonstration of the cuttable and recyclable WPT sheet, including fabrication, reshaping, wireless powering, dissolution, and refabrication.*

Figure 7 shows the recyclability of the galinstan. 98% of the injected galinstan was recovered by dissolving the PVA channels in water. The loss due to mixing into dissolved PVA, adhesion to the container, or dispersion in water was negligible. During four dissolution–refabrication cycles, the volume resistivity of the recovered galinstan remained stable at $0.32 \pm 0.01$ mΩ·mm. The contact resistance between the galinstan and pin headers remained stable at $11.7 \pm 1.3$ mΩ. These results confirm that the dissolution and refabrication processes do not degrade the electrical properties of galinstan, thereby demonstrating excellent material recyclability.

Finally, a cuttable and recyclable WPT sheet was demonstrated using a 4 × 4 coil array prototype (Figure 8). The sheet was fabricated by 3D printing and injection, and the power transfer area was reshaped by cutting. Multiple LED modules were successfully powered by the sheet. The control boards and galinstan were recovered by dissolving the sheet in water. Another WPT sheet was refabricated with the recovered components and exhibited WPT performance. Overall, the proposed WPT sheet enabled a complete cycle of fabrication, cutting, operation, recycling, and refabrication. This study advances circular flexible electronics that users can easily reshape, disassemble, and refabricate on demand.

## CONCLUSIONS

A recyclable and cuttable WPT sheet was developed, enabled by H-tree wiring pattern and 3D-printed PVA channels filled with galinstan. The channel geometry was optimized to maximize power transfer efficiency with mechanical flexibility and cuttability. The sheet retained its durability under repeated bending and maintained stable electrical properties through multiple recycling cycles. These results demonstrate that the sheet can be reshaped and refabricated while preserving its WPT performance. The proposed sheet provides new opportunities for user-configurable and recyclable wireless power systems that can be integrated into future living environments and IoT infrastructure.

## ACKNOWLEDGEMENTS

This work was partly supported by JST PRESTO JPMJPR2515, JSPS KAKEN 22K21343, and JST ASPIRE JPMJAP2401.

## CONTACT

T. Sato  tel: +81-80-2185-3561
e-mail: machotakashi-satou@aist.go.jp